\documentclass[a4paper,10pt]{article}
\pdfoutput=1

\usepackage[T1]{fontenc}
\usepackage[utf8]{inputenc}
\usepackage{lmodern}
\usepackage{microtype}

\usepackage{indentfirst}

\usepackage{amsmath}
\usepackage{ctable}

\newcommand{\corE}{black}

\usepackage[%
        pdftex,
        colorlinks=true,
        breaklinks=true,
        bookmarks=true,
        bookmarksnumbered=true,
        bookmarksopen=true,%
        plainpages=false,
        breaklinks=true,
        anchorcolor=\corE,
        citebordercolor=\corE,
        citecolor=\corE,
        linkcolor=\corE,
        linkbordercolor=\corE,
        filebordercolor=\corE,
        menubordercolor=\corE,
        urlcolor=\corE,
        urlbordercolor=\corE,
        pdftitle={Geometry-Topology Duality in complex porous networks},
        pdfauthor={André Rafael Cunha},
        pdfsubject={Porous networks},
        pdfkeywords={Porous media; Porous networks; Permeability},
        pdflang=English
]{hyperref}

\begin{document}

% % % % % % % % % % % % % % % % % % % % % % % % % % % % % % % % % % % % % % % % 
% % % % % % % % % % % % % % % % % % % % % % % % % % % % % % % % % % % % % % % % 
% % % % % % % % % % % % % % % % % % % % % % % % % % % % % % % % % % % % % % % % 
% 
% 
% 

\title{Geometry-Topology Duality in complex porous networks%
\footnote{Research supported by Brazilian agencies CAPES, CNPq and FAPESC, and Petrobras.}%
}

\author{%
    Andr\'e Rafael Cunha\footnote{A. R. Cunha. Porous Media and Thermophysical Properties Laboratory, Federal University of Santa Catarina, SC, Brazil. \url{andre.cunha@posgrad.ufsc.br}.}
    \and
    Celso Peres Fernandes
    \and
    Lu\'is Orlando Emerich dos Santos
}

\maketitle

\begin{abstract}
We explore the experimental observation that complex networks of porous media exhibit the property that the porous coordination number is proportional to its size.
Based in this geometry-topology duality we developed an analytical approach to describe the permeability transport property of the material.
That results was compared with 2D networks simulations.
And we found a correlation coefficient greater than 95\%.

\medskip

Porous media.
Transport properties.

\medskip


\end{abstract}

% % % % % % % % % % % % % % % % % % % % % % % % % % % % % % % % % % % % % % % % 
% % % % % % % % % % % % % % % % % % % % % % % % % % % % % % % % % % % % % % % % 
% % % % % % % % % % % % % % % % % % % % % % % % % % % % % % % % % % % % % % % % 
% 
% 
% 

\section{Introduction}

A useful way to modeling porous media structure is to consider the pore space as a network formed by \textit{pores}, larger spaces that store fluid, and \textit{throats}, which restrict the flow while performing the communication between the pores \cite{Dullien1979,Dagan1989}.
Under this view, two quantities are relevant for the displacement of matter: the radius of the pore, and the number of throats that leave that pore.
The first is of geometric nature, and the second, topological.
The number of throats connected into a specific pore is called the \textit{coordination number} of that pore.

The representation ways of the porous medium by network are related to the development of computation, 
since a network is formed by many constituents;
and due to imaging techniques, which can provide information about the material.
In the 1950s, some authors used to circumvent the problem of excessive calculations by means of electromechanical analogies 
\cite{Bruce1943,Scheidegger1963,Owen1952,Sahimi1993,Fatt1956a,Fatt1956b,Fatt1956c}.
At an intermediate stage,
the 2D imaging techniques allowed the introduction of images to the simulation.
However a 2D image is not able to adequately represent porous space connectivity \cite{ChatzisDullien1977,VanMarcke2010}).
Therefore, criteria were developed to generate new random networks from statistical image information,
which are superimposed to build a 3D volume where the phenomenon is simulated.
And even if higher-order statistics \cite{Okabe2005} or multiscalar schemes \cite{Fernandes1996,Graham2012,Tsakiroglou2012,Ichikawa2012} are considered, 
the generated volume does not properly express the real pore space.
The advent of the X-ray microtomography technique in porous media research in the 1980s \cite{Vinegar1987,Dunsmuir1991,BryantBlunt1992,Landisa2010,Russ2011,Sun2012,Mariam2010}
made possible the observation of the real porous space complexity.
A phenomenon can be simulated in a $\mu$-CT image using a conventional numerical method \cite{VanMarcke2010}, 
but this procedure is computationally time consuming.
Therefore some simplification of the image is still attempted,
and several methods return to the idea of a network \cite{AlKharusi2007},
but in this case, the spatial configuration is represented with greater authenticity.

When geometrical shapes are assigned to the pore space, it is called a \textit{pore-throat network} or morphological network,
where the phenomenon is described by the conservation laws, i.e., the \textit{continuum models}.
In \textit{random networks}, the phenomenon is approached by statistical physical theories 
based on results of theories of percolation, renormalization, fractals and cellular automata, for example.
They are \textit{discrete models} \cite{Sahimi1993}.
In this work, we apply the Maximum Ball Algorithm \cite{HuDongPRE2009} to the microtomographic image.
The result is a network of spherical pores and cylindrical throats.
In some cases, the simulation based on the discretization of the motion equations is summarized by a linear system \cite{VanMarcke2010}.

\section{Statistical properties of complex porous networks}

\begin{figure}
\centering
\caption{PSD of the sandstone samples.}
\includegraphics[keepaspectratio=true,width=0.85\textwidth]{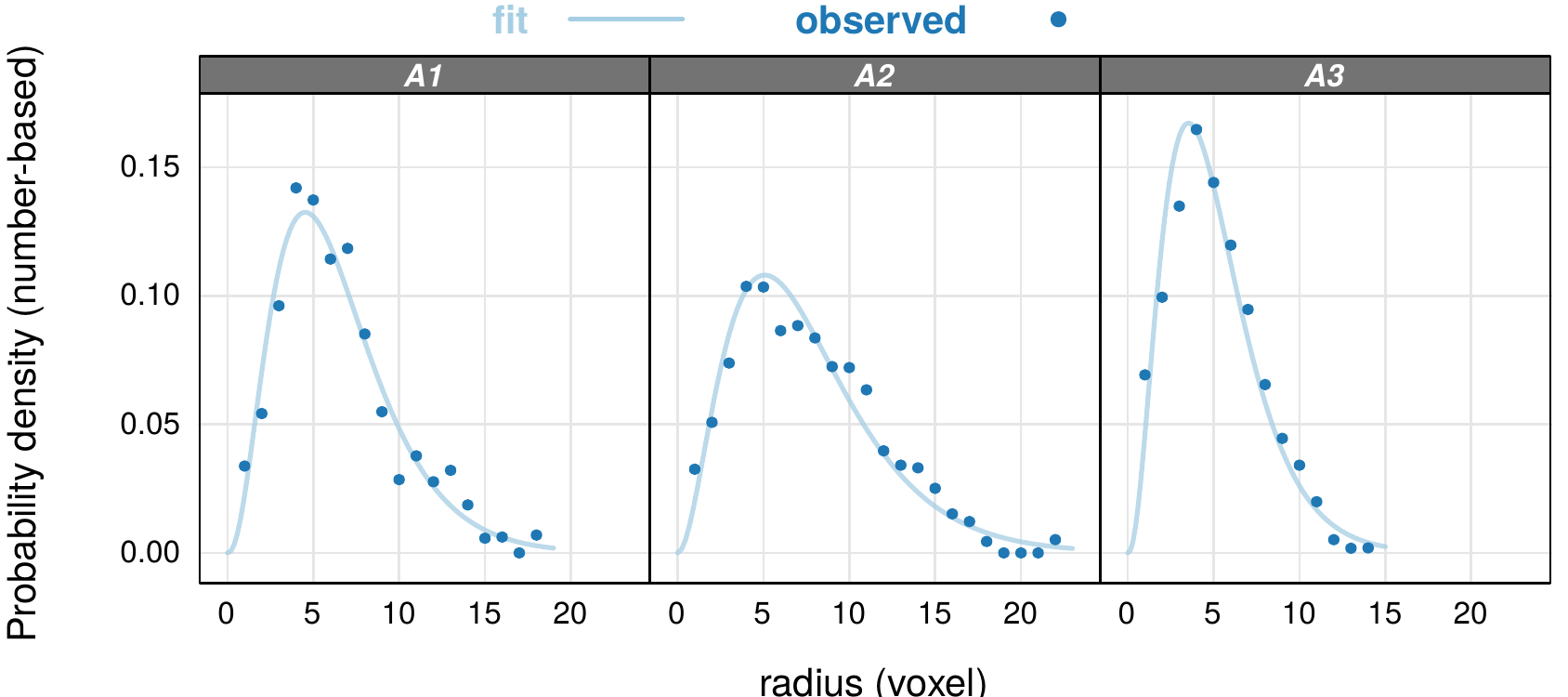}
\label{Graf:DTPAren}
\end{figure}

It is observed that the Pore Size Distribution (PSD) for the sandstone samples can be approximated by a gamma distribution (Fig. \ref{Graf:DTPAren})
\begin{align}
    R(r) &= \frac{1}{\Gamma(\alpha)} \beta^{\alpha} r^{\alpha-1} e^{-\beta r} \quad, \quad r \geq 0 \quad.
    \label{Eq:rGama}
\end{align}
where $\alpha$ and $\beta$ are the parameters of the distribution, and $\Gamma(x)$ is the gamma function.

\begin{figure}
\centering
\includegraphics[keepaspectratio=true,width=0.8\textwidth]{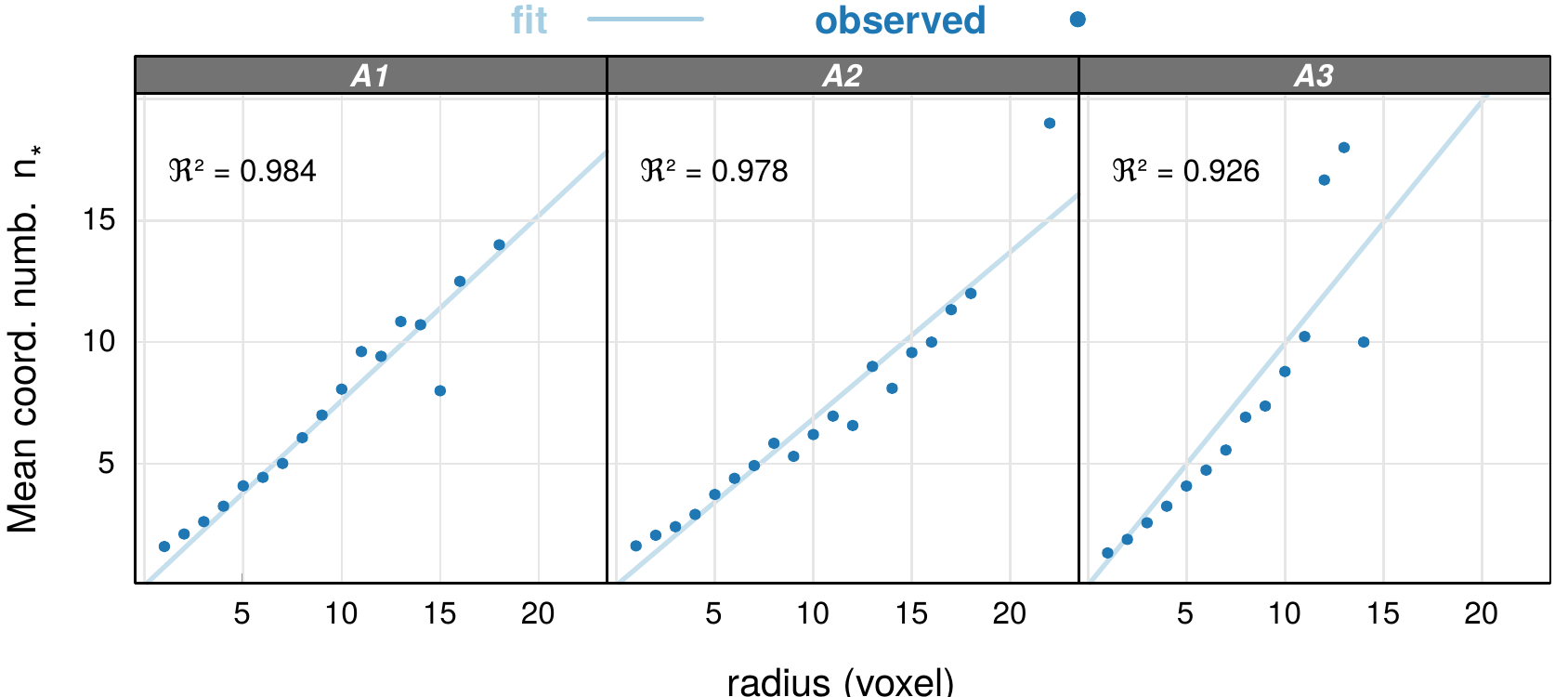}
\caption{Correlation between $n_{*}$ and $r$.}
\label{Graf:CorrelAren}
\end{figure}

The existence of a linear correlation between the mean coordination number $n_{*}$ of a pore and its radius $r$ (Fig.~\ref{Graf:CorrelAren}) is also observed, which characterizes a geometry-topology duality \cite{Flegg1974}.
Mathematically:
\begin{align}
    n_{*} &\sim r \quad, \\
    n_{*} &= a r + b \quad, 
\end{align}
where $b=0$ since  a pore with $r=0$ does not exist and implies no connected throat.
Then,
\begin{align}
    n_{*} &= a r \quad.    \label{Eq:Correlacao}
\end{align}

The observed correlation implies that one can express $N_{*}(n_{*})$ in terms of $R(r)$ \cite{Kay2005,Papoulis1991}:
{\allowdisplaybreaks%
\begin{align}
    N_{*}(n_{*}) &= \frac{1}{a}\ R \left( \frac{n_{*}}{a} \right)  \quad.
\end{align}}%
which means that once the geometry of the network is known, the topology is also;
and the reverse is also true.

\subsection{A phenomenological connectivity}

Faced with a flow in the pore-throat network, two quantities are relevant for the mass displacement: the pore radius, and its coordination number.
Highlighting the interaction between these entities of spatial configuration,
we define the connectivity function $\xi$  
\cite{Vasconcelos1998,Bernabe2016,Glover2010,Montaron2009} 
as
\begin{align}
    \xi(n_{*},r) &\sim N_{*}(n_{*})\ R(r) \quad. \label{Eq:ConectividadeMedia}
\end{align}
By eq. \ref{Eq:Correlacao},
\begin{align}
    \xi(r) &\sim R(r)^2 \quad. \label{Eq:XiSimRR}
\end{align}
The function $\xi$ can be characterized by only one of the variables $n_{*}$ or $r$.
The radius $r$ was chosen because it can be measured by different techniques.
Normalizing the equation%
\footnote{The normalization condition requires that: $\alpha>\frac{1}{2}$ e $\beta>0$.}%
,
\begin{align}
    \xi(r) &= \frac{2^{2 \alpha -1} \beta  \Gamma (\alpha )^2}{\Gamma (2 \alpha -1)}\ R(r)^2 \quad. \label{Eq:Normalizacao}
\end{align}

\subsection{Application to permeability}

The permeability $k$ is given by \cite{Scheidegger1963,Nutting1930,Tiab2004}:
\begin{align}
    k &= - \frac{\eta L Q}{A \left( p_\text{out} - p_\text{in} \right)} \quad ,       \label{Eq:k}
\end{align}
where $\eta$ is the viscosity of the fluid, $L$ is the length of the material, $A$ is the area of the section, $p_\text{in}$ and $p_\text{out}$ are the pressures applied at the inlet and outlet ends, respectively, and $Q$ is the flow through $A$. 
The flow $Q$ is given by
\begin{align*}
    Q &= - \frac{1}{\Omega} \left( p_\text{out} - p_\text{in} \right) \quad ,
\end{align*}
where $\Omega$ is he hydraulic resistance of the porous medium.
Then, the eq. (\ref{Eq:k}) becomes
\begin{align}
    k &= \frac{L\eta}{A\Omega} \quad .       \label{Eq:k_novo}
\end{align}
On the right side, $L$, $A$ and $\Omega$ are all macroscopic ($\eta$ is a fluid property).
But the hydraulic resistance $\Omega$ is affected by the microscopic characteristics of the porous space.
We then consider $\Omega$ as a mean of microscopic informations.

Strictly speaking, the hydraulic resistance of a cell in the pore-throat network has two parts:
\begin{align*}
\Omega = \Omega_p + \Omega_g \quad ,
\end{align*}
where $\Omega_p$ is the contribution due to the spherical pores,
and $\Omega_g$ is due to cylindrical throats.
The connectivity $\xi(r)$ explicitly considers only the radius $r$ of the pores and the average number of throats that depart from that pore,
but not the geometry of those links.
However the Maximum Ball Algorithm establishes a relation between the geometries of the radius and its connected throat \cite{HuDongPRE2009}. 
Then we can rewrite%
% 
% % \footnote{A contribuição das gargantas Por essa correlação, incorpora-se a contribuição $\Omega_g$ à parcela dos poros $\Omega_p$ por uma constante de ajuste $\cteBM$, que é calibrada por comparação com os valores experimentais.\alerta{}}
% 
, 
\begin{align}
    \Omega  &= \Omega_p\ (1+\tau) \quad , \nonumber \\
    \Omega  &\propto \Omega_p \quad . \nonumber 
\end{align}%
Since the objective of this work is to demonstrate the existence of a correlation,
we write, without loss of generality,
\begin{align}
    \Omega  &= \Omega_p \quad , \nonumber
\end{align}
therefore
\begin{align}
    \Omega  &= \frac{81 \eta}{\pi r^3} \quad . \label{Eq:Omega}
\end{align}

At this point the connectivity function $\xi$ is used to weight an expected value of $r^3$ in eq. (\ref{Eq:Omega})
{\allowdisplaybreaks%
\begin{align*}
    \langle r^3\rangle_\xi &= \int_{0}^\infty \xi(r)\ r^3\ dr \quad,    \\
    \langle r^3\rangle_\xi &= \frac{1}{4} \alpha  \left(4 \alpha ^2-1\right) \beta ^3  \quad.
\end{align*}}%\usepackage{

Replacing eq. (\ref{Eq:Omega}) in eq. (\ref{Eq:k_novo}),
\begin{align}
    k &= \frac{\pi L}{81 A} \; \frac{4}{\alpha  \left(4 \alpha ^2-1\right) \beta ^3}  \quad .       \label{Eq:k_final}
\end{align}

\section{Materials}

We assign values to $\alpha$ e $\beta$ in such a way to span a wide range of porous size and consequently of permeability (Tabs. \ref{Tab:ParamA} and \ref{Tab:ParamB}).
An observation: $L$ can be considered dimensionless since we are interesting only in the correlation.

\begin{table}[!h]
\caption{The first 30 networks' parameters e permeability values.}
\label{Tab:ParamA}
\centering %\footnotesize
\begin{tabular}{cccc|cc}                                                     \FL
   Network &   $\alpha$ &    $\beta$ &   $L$  &  $k_{s}$     &   $k_{a}$    \ML 
    d01    &   {4.30}   &     {96}   &   {15} &  {5,04E-10}  &   {5,04E-10} \NN 
    d02    &   {4.90}   &     {12}   &   {49} &  {3,63E-12}  &   {3,63E-12} \NN
    d03    &   {2.20}   &     {92}   &   {9}  &  {6,52E-21}  &   {6,52E-21} \NN
    d04    &   {4.15}   &     {38}   &   {33} &  {2,11E-15}  &   {2,11E-15} \NN
    d05    &   {4.25}   &     {108}  &   {12} &  {2,41E-20}  &   {2,41E-20} \NN
    d06    &   {2.95}   &     {118}  &   {6}  &  {4,72E-17}  &   {4,72E-17} \NN
    d07    &   {4.45}   &     {70}   &   {25} &  {8,63E-18}  &   {8,63E-18} \NN
    d08    &   {3.30}   &     {122}  &   {6}  &  {2,75E-20}  &   {2,75E-20} \NN
    d09    &   {2.20}   &     {20}   &   {21} &  {1,77E-17}  &   {1,77E-17} \NN
    d10    &   {2.15}   &     {48}   &   {16} &  {2,48E-10}  &   {2,48E-10} \NN
    d11    &   {3.35}   &     {30}   &   {29} &  {1,52E-17}  &   {1,52E-17} \NN
    d12    &   {3.55}   &     {74}   &   {19} &  {4,73E-18}  &   {4,73E-18} \NN
    d13    &   {4.75}   &     {52}   &   {33} &  {1,34E-15}  &   {1,34E-15} \NN
    d14    &   {3.80}   &     {126}  &   {5}  &  {7,57E-19}  &   {7,57E-19} \NN
    d15    &   {3.40}   &     {38}   &   {27} &  {1,26E-15}  &   {1,26E-15} \NN
    d16    &   {3.95}   &     {40}   &   {31} &  {1,50E-15}  &   {1,50E-15} \NN
    d17    &   {4.00}   &     {122}  &   {7}  &  {1,45E-16}  &   {1,45E-16} \NN
    d18    &   {2.80}   &     {14}   &   {28} &  {2,83E-15}  &   {2,83E-15} \NN
    d19    &   {3.55}   &     {130}  &   {4}  &  {2,38E-20}  &   {2,38E-20} \NN
    d20    &   {2.80}   &     {128}  &   {4}  &  {5,59E-17}  &   {5,59E-17} \NN
    d21    &   {2.30}   &     {16}   &   {22} &  {3,44E-15}  &   {3,44E-15} \NN
    d22    &   {4.85}   &     {126}  &   {7}  &  {3,51E-20}  &   {3,51E-20} \NN
    d23    &   {3.75}   &     {114}  &   {9}  &  {1,67E-10}  &   {1,67E-10} \NN
    d24    &   {2.05}   &     {82}   &   {10} &  {4,39E-15}  &   {4,39E-15} \NN
    d25    &   {3.70}   &     {22}   &   {34} &  {4,06E-17}  &   {4,06E-17} \NN
    d26    &   {3.60}   &     {70}   &   {20} &  {4,99E-16}  &   {4,99E-16} \NN
    d27    &   {2.80}   &     {106}  &   {8}  &  {2,22E-14}  &   {2,22E-14} \NN
    d28    &   {2.15}   &     {16}   &   {21} &  {4,82E-09}  &   {4,82E-09} \NN
    d29    &   {3.35}   &     {88}   &   {14} &  {7,12E-13}  &   {7,12E-13} \NN
    d30    &   {2.40}   &     {124}  &   {4}  &  {1,36E-14}  &   {1,36E-14} \LL
\end{tabular}
\end{table}

\clearpage

\begin{table}[!h]
\caption{The last 30 networks' parameters e permeability values.}
\label{Tab:ParamB}
\centering %\footnotesize
\begin{tabular}{cccc|cc}                                                     \FL
   Network &   $\alpha$ &    $\beta$ &   $L$  &  $k_{s}$     &   $k_{a}$    \ML 
    d30    &   {2.40}   &     {124}  &   {4}  &  {1,36E-14}  &   {1,36E-14} \NN
    d31    &   {2.90}   &     {32}   &   {24} &  {5,04E-10}  &   {5,04E-10} \NN
    d32    &   {4.45}   &     {24}   &   {40} &  {3,63E-12}  &   {3,63E-12} \NN
    d33    &   {4.30}   &     {88}   &   {19} &  {6,52E-21}  &   {6,52E-21} \NN
    d34    &   {4.60}   &     {16}   &   {44} &  {2,11E-15}  &   {2,11E-15} \NN
    d35    &   {5.00}   &     {60}   &   {31} &  {2,41E-20}  &   {2,41E-20} \NN
    d36    &   {4.75}   &     {40}   &   {37} &  {4,72E-17}  &   {4,72E-17} \NN
    d37    &   {4.80}   &     {68}   &   {28} &  {8,63E-18}  &   {8,63E-18} \NN
    d38    &   {4.75}   &     {10}   &   {48} &  {2,75E-20}  &   {2,75E-20} \NN
    d39    &   {4.50}   &     {106}  &   {13} &  {1,77E-17}  &   {1,77E-17} \NN
    d40    &   {3.95}   &     {78}   &   {20} &  {2,48E-10}  &   {2,48E-10} \NN
    d41    &   {2.85}   &     {74}   &   {16} &  {1,52E-17}  &   {1,52E-17} \NN
    d42    &   {3.05}   &     {128}  &   {4}  &  {4,73E-18}  &   {4,73E-18} \NN
    d43    &   {3.00}   &     {24}   &   {28} &  {1,34E-15}  &   {1,34E-15} \NN
    d44    &   {2.10}   &     {72}   &   {12} &  {7,57E-19}  &   {7,57E-19} \NN
    d45    &   {4.70}   &     {114}  &   {11} &  {1,26E-15}  &   {1,26E-15} \NN
    d46    &   {4.45}   &     {112}  &   {11} &  {1,50E-15}  &   {1,50E-15} \NN
    d47    &   {2.70}   &     {40}   &   {21} &  {1,45E-16}  &   {1,45E-16} \NN
    d48    &   {2.75}   &     {116}  &   {6}  &  {2,83E-15}  &   {2,83E-15} \NN
    d49    &   {4.70}   &     {100}  &   {15} &  {2,38E-20}  &   {2,38E-20} \NN
    d50    &   {4.65}   &     {130}  &   {5}  &  {5,59E-17}  &   {5,59E-17} \NN
    d51    &   {3.50}   &     {78}   &   {18} &  {3,44E-15}  &   {3,44E-15} \NN
    d52    &   {2.95}   &     {124}  &   {5}  &  {3,51E-20}  &   {3,51E-20} \NN
    d53    &   {2.45}   &     {72}   &   {14} &  {1,67E-10}  &   {1,67E-10} \NN
    d54    &   {3.20}   &     {58}   &   {21} &  {4,39E-15}  &   {4,39E-15} \NN
    d55    &   {4.45}   &     {66}   &   {27} &  {4,06E-17}  &   {4,06E-17} \NN
    d56    &   {3.60}   &     {10}   &   {37} &  {4,99E-16}  &   {4,99E-16} \NN
    d57    &   {4.30}   &     {100}  &   {15} &  {2,22E-14}  &   {2,22E-14} \NN
    d58    &   {3.45}   &     {74}   &   {19} &  {4,82E-09}  &   {4,82E-09} \NN
    d59    &   {2.60}   &     {22}   &   {24} &  {7,12E-13}  &   {7,12E-13} \NN
    d60    &   {4.80}   &     {46}   &   {36} &  {1,36E-14}  &   {1,36E-14} \LL
\end{tabular}
\end{table}

\clearpage

\section{Results}

We proceed the comparison of the results from eq. (\ref{Eq:k_final}) with those obtained from 2D simulations, based on the solution of the discretized mass conservation equation $\sum Q_{ij}=0$ for the the system \cite{HuDongPRE2009}.
% . %Fig. \ref{Fig:Network01} shows the pressure field calculated for the d01 sample.

% 
% \begin{figure}
% \centering
% \includegraphics[keepaspectratio=true,width=0.8\textwidth]{fig_network_n01.pdf}
% \caption{Pressure field obtained by the simulation for d01 sample.}
% \label{Fig:Network01}
% \end{figure}%

\begin{figure}[!h]
\centering
\includegraphics[keepaspectratio=true,width=0.8\textwidth]{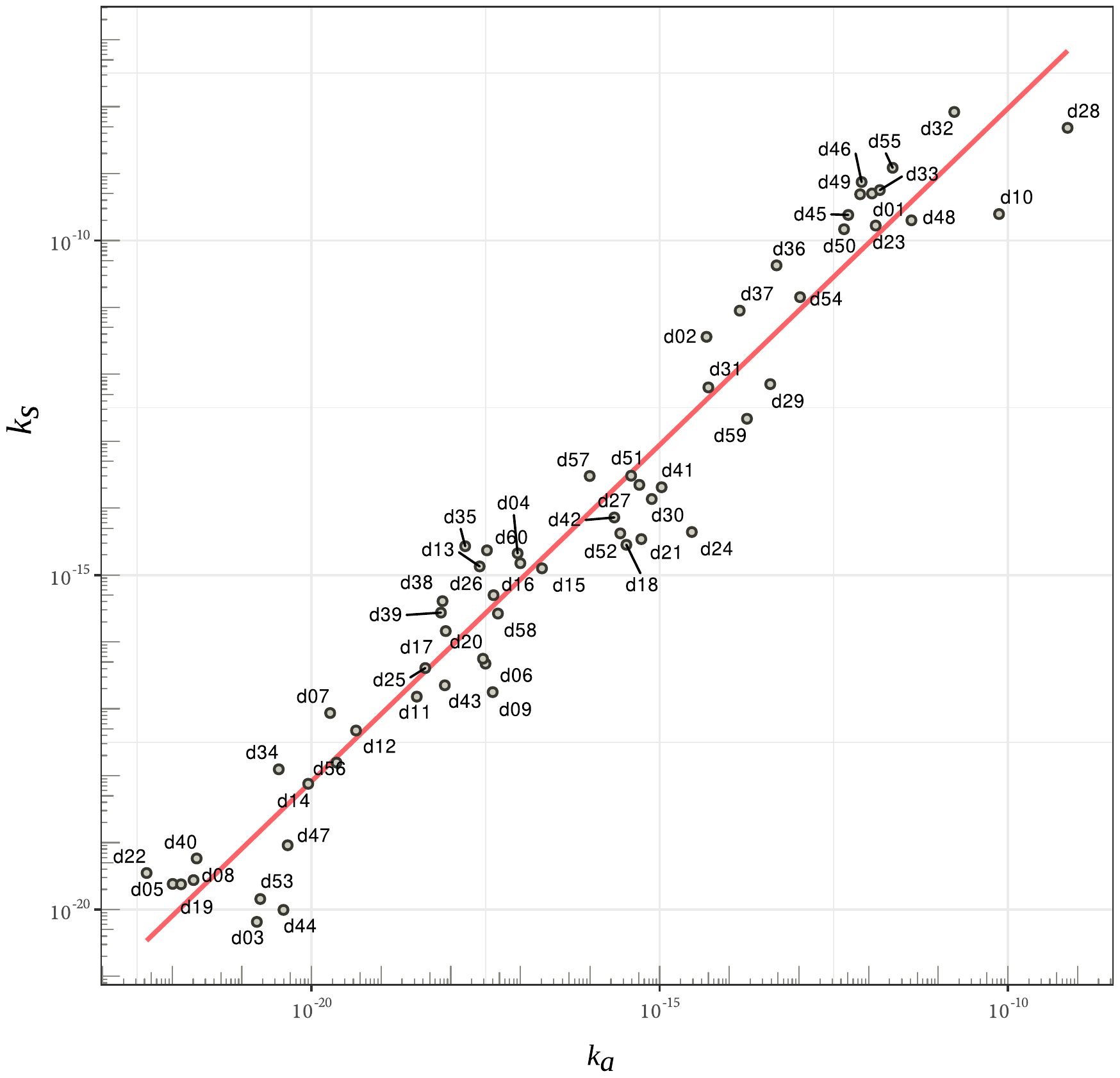}
\caption{Comparison between $k_s$ and $k_a$.}
\label{Graf:Correl}
\end{figure}

The second part of Tabs. \ref{Tab:ParamA} and \ref{Tab:ParamB} shows the permeability values:
$k_s$ is obtained from the simulation
and $k_a$ from eq. \ref{Eq:k_final}.
The comparison between both are exposed in Fig. \ref{Graf:Correl}.
We observe a correlation coeficient r-squared 95.05\%.

% \clearpage

\section{Conclusions}

The existence of a correlation between the mean coordination number of a pore and its size -- the geometry-topology duality -- in complex porous networks allow us to conceive a connectivity density function.
Using that function to weight the porous transport property, we reached the values of permeability of the material.
Those values was compared with other obtained from 2D simulations.

As result we observe the existence of a correlation of 95.05\% between them,
which means that the geometry-topology duality can serve as an new path to describe complex porous networks properties.

% Non-BibTeX users please use

\end{document}